\documentclass[aps,pre,groupedaddress,onecolumn]{revtex4-2}

\usepackage{color}

\usepackage[utf8]{inputenc}
\usepackage[english]{babel}
\usepackage{amsmath,graphicx,enumerate}

\usepackage{amsmath,amsfonts}
\usepackage{graphicx}
\usepackage{booktabs,caption}
\usepackage{epsfig}
\usepackage{wrapfig}
\usepackage{subcaption}
\usepackage{hyperref}

\usepackage{amsthm}

  \newcommand{\av}[1]{\left\langle#1\right\rangle}
  \newcommand{\cbr}[1]{\left(#1\right)}

  \newcommand{\sbr}[1]{\left[#1\right]}

\begin{document}

\title{Ergodic observables in non-ergodic systems: the example of the harmonic 
chain}


\author{Marco Baldovin}

\affiliation{Institute for Complex Systems - CNR, P.le Aldo Moro 2, 00185, Rome, Italy}
            
 \affiliation{Université Paris-Saclay, CNRS, LPTMS,530 Rue André Rivière, 91405, Orsay, France}

\author{Raffaele Marino}%
\affiliation{Dipartimento di Fisica e Astronomia, 
Universit\'{a} degli Studi di Firenze, Via Giovanni Sansone 1, 50019, Sesto Fiorentino, Italy}

 \author{Angelo Vulpiani}%

\affiliation{Dipartimento di Fisica, Sapienza Universit\'{a} di 
Roma, P.le Aldo Moro 5, 00185, Rome, Italy}

\date{\today}

\begin{abstract}
In the framework of statistical mechanics the properties of
macroscopic systems are deduced starting from the laws of their microscopic
dynamics. One of the key assumptions in this procedure is the ergodic property, 
namely the equivalence between time averages and ensemble averages.
This property can be proved only for a limited number of systems; however,
as proved by Khinchin~\cite{aleksandr1949mathematical}, weak forms of it hold even in systems that
are not ergodic at the microscopic scale, provided that extensive observables are considered.

Here we show in a pedagogical way the
validity of the ergodic hypothesis, at a \textit{practical} level,
in the paradigmatic case of a chain of harmonic oscillators. By
using analytical results and numerical computations, we provide evidence that
this non-chaotic integrable system shows ergodic behavior in the
limit of many degrees of freedom. In particular, the
Maxwell-Boltzmann distribution turns out to fairly describe the
statistics of the single particle velocity. A study of the
typical time-scales for relaxation is also provided.
\end{abstract}

\maketitle








\section{Introduction}
\label{sec::intr}

Since the seminal works by Maxwell, Boltzmann and Gibbs, statistical mechanics 
has been conceived as a link between the microscopic world of atoms and 
molecules and the macroscopic one where everyday phenomena are 
observed~\cite{huang2009introduction}. The same physical system can be 
described, in the former, by an enormous number of degrees of freedom $N$ (of 
the same order of the Avogadro number) or, in the latter, in terms of just a few 
thermodynamics quantities. Statistical mechanics is able to describe in a 
precise way the behavior of these macroscopic observables, by exploiting the 
knowledge of the laws for the microscopic dynamics and classical results from 
probability theory. Paradigmatic examples of this success are, for instance, the
possibility to describe the probability distribution of the single-particle velocity
in an ideal gas~\cite{huang2009introduction, peliti2003}, as well as  the detailed behavior
of phase transitions~\cite{ehrenfest1990conceptual, di2015statistical} and
critical phenomena~\cite{ma1985statistical, caraccciolo2021criticality}.
In some cases (Bose-Einstein condensation~\cite{anderson1995}, absolute negative temperature
systems~\cite{ramsey1956thermodynamics, braun2013, baldovin2021statistical}) the results
of statistical mechanics were able to predict states of the matter that were never been observed before.

In spite of the above achievements, a complete consensus about the actual reasons for such a success 
has not been yet reached within the statistical mechanics community. The main 
source of disagreement is the so-called ``ergodic hypothesis'', stating that time 
averages (the ones actually measured in physics experiments) can be computed as 
ensemble averages (the ones appearing in statistical mechanics calculations). 
Specifically, a system is called ergodic when the value of the time average of 
any observable is the same as the one obtained by taking the average over the 
energy surface, using the microcanonical distribution~\cite{lebowitz1973modern}. It is worth mentioning that, from a mathematical 
point of view, ergodicity holds only for a small amount of physical systems: the 
KAM theorem~\cite{arnold2009proof, arnol2013mathematical, arnold1988dynamical} 
establishes that, strictly speaking, non-trivial dynamics cannot be ergodic. 
Nonetheless, the ergodic hypothesis happens to work extremely well also for 
non-ergodic systems. It provides results in perfect agreement with the numerical and experimental observations, as seen in a wealth of physical situations~\cite{livi1987,baldovin2021, cocciaglia2022thermalization}.

Different explanations for this behavior have been provided. Without going into 
the details of the controversy, three main points of view can be identified: (i) 
the ``classical'' school based on the seminal works by Boltzmann and the important 
contribution of Khinchin, where the main role is played by the presence of many 
degrees of freedom in the considered systems 
~\cite{aleksandr1949mathematical, schrodinger1944statistical, darrigol2018, boltzmann2022lectures, 
boltzmann2012theoretical, castiglione2008}; (ii) those, like the \textit{Prigogine} school, who recognize in the chaotic 
nature of the microscopic evolution  the dominant ingredient~\cite{prigogine1978time, prigogine2017non, 
prigogine1996laws, lando2023thermalization}; (iii) the \textit{maximum entropy} point of view, which does 
not consider statistical mechanics as a physical theory but as an inference 
methodology based on incomplete information~\cite{jaynes1982rationale, 
uffink1995can, jaynes1957information, jaynes2003probability}.

The main aim of the present contribution is to clarify, at a pedagogical level, 
how ergodicity manifests itself  for some relevant degrees of freedom, in 
non-ergodic systems. We say that ergodicity occurs ``at a practical level''. 
To this end, a classical chain of $N$ coupled harmonic oscillators turns out to 
be an excellent case study: being an integrable system, it cannot be suspected 
of being chaotic; still, ``practical'' ergodicity is recovered for relevant 
observables, in the limit of $N\gg1$.  We believe that this kind of analysis 
supports the traditional point of view of Boltzmann, which identifies the large 
number of degrees of freedom as the reason for the occurrence of ergodic 
behavior for physically relevant observables. Of course, these conclusions are 
not new. In the works of Khinchin (and then Mazur and van der 
Lynden)~\cite{aleksandr1949mathematical, halmos2017lectures, van1967asymptotic, 
mazur1963asymptotic, casetti1997fermi} it is rigorously shown that the ergodic 
hypothesis holds for observables that are computed as an average over a finite 
fraction of the degrees of freedom, in the limit of $N \gg 1$. Specifically, if 
we limit our interest to this particular (but non-trivial) class of observables, 
the ergodic hypothesis holds for almost all initial conditions (but for a set 
whose probability goes to zero for $N \to \infty$), within arbitrary accuracy. 
In addition, several numerical results for weakly non-linear systems 
~\cite{chakraborty2001dynamics, arnold1978white, arnold2012stochastic}, as well 
as integrable systems~\cite{arnol2013mathematical, goldstein2002classical}, 
present  strong indications of the poor role of chaotic behaviour, implying the 
dominant relevance of the many degrees of freedom. Still, we think it may be 
useful, at least from a pedagogical point of view, to analyze an explicit 
example where analytical calculations can be made (to some extent), without 
losing physical intuition about the model.

The rest of this paper is organized as follows. In Section~\ref{sec::model} we 
briefly recall basic facts about the chosen model, to fix the notation and 
introduce some formulae that will be useful in the following. 
Section~\ref{sec::deterministic} contains the main result of the paper. We present an explicit calculation of the empirical distribution of the single-particle momentum, given a system starting from out-of-equilibrium initial conditions. We show that in this case the Maxwell-Boltzmann distribution is an excellent approximation in the $N\to \infty$ limit. Section~\ref{sec::times} is devoted to an analysis 
of the typical times at which the described ergodic behavior is expected to be 
observed; a comparison with a noisy version of the model (which is ergodic by 
definition) is also provided. In Section~\ref{sec::conclusion} we draw our final 
considerations.

\section{Model}
\label{sec::model}

We are interested in the dynamics of a one-dimensional chain of $N$ classical 
harmonic oscillators of mass $m$.
The state of the system is described by the canonical coordinates $\{q_j(t), 
p_j(t)\}$ with $j=1,..,N$; here $p_j(t)$ identifies the momentum of the $j$-th 
oscillator at time $t$, while $q_j(t)$ represents its position. The $j$-th and 
the $(j+1)$-th particles of the chain interact through a linear force of 
intensity $\kappa|q_{j+1}-q_j|$, where $\kappa$ is the elastic constant. We will 
assume that the first and the last oscillator of the chain are coupled to 
virtual particles at rest, with infinite inertia (the walls), i.e.  $q_{0}\equiv q_{N+1} \equiv 0$.

The  Hamiltonian of the model reads therefore
\begin{equation}
    \mathcal{H}(\mathbf{q},\mathbf{p})=\sum_{j=0}^N \frac{p_j^2}{2 m} + 
\sum_{j=0}^{N} \frac{m \omega_0^2 }{2}(q_{j+1} - q_{j})^2,
    \label{H:integrable}
\end{equation}
where $\omega_0=\sqrt{\frac{\kappa}{m}}$.

Such a system is integrable and, therefore, trivially non-ergodic. This can 
be easily seen by considering the normal modes of the chain, i.e. the set of 
canonical coordinates

\begin{subequations}
\label{eq:modes}
\begin{equation}
\label{eq:modes_q}
Q_k=\sqrt{\frac{2}{N+1}}\sum_{j=1}^N\, q_j \sin\cbr{\frac{j k \pi}{N+1}}
\end{equation}
\begin{equation}
\label{eq:modes_p}
P_k=\sqrt{\frac{2}{N+1}}\sum_{j=1}^N\, p_j \sin\cbr{\frac{j k \pi}{N+1}}\,,
\end{equation}
\end{subequations}
with $k=1, ..., N$. Indeed, by rewriting the Hamiltonian in terms of these new 
canonical coordinates one gets
\begin{equation}
    \mathcal{H}(\mathbf{Q},\mathbf{P})=\frac{1}{2}\sum_{k=1}^N 
\cbr{\frac{P_k^2}{m} + \omega_k^2 Q_k^2 },
    \label{H:normal}
\end{equation}
where the frequencies of the normal modes are given by
\begin{equation}
\label{eq:omega_m}
\omega_k=2 \omega_0 \sin \cbr{\frac{\pi k}{2N +2}}\,.
\end{equation}
In other words, the system can be mapped into a collection of independent 
harmonic oscillators with characteristic frequencies $\{\omega_k\}$. This system 
is clearly non-ergodic, as it admits $N$ integrals of motion, namely the 
energies
$$
E_k=\frac{1}{2} \cbr{\frac{P_k^2}{m} + \omega_k^2 Q_k^2 }
$$
associated to the normal modes. 

In spite of its apparent simplicity, the  above  system allows the investigation 
of some nontrivial aspects of the ergodic hypothesis, and helps clarifying the 
physical meaning of this assumption.

\section{Ergodic behavior of the momenta}
\label{sec::deterministic}
In this section we analyze the statistics of the single-particle momenta of the 
chain. We aim to show that they approximately follow a Maxwell-Boltzmann 
distribution
\begin{equation}
\label{eq:mb}
\mathcal{P}_{MB}(p)=\sqrt{\frac{\beta}{2\pi m}}e^{-\beta p^2/2m}
\end{equation}
in the limit of large $N$, where $\beta$ is the inverse temperature of the 
system. With the chosen initial conditions, $\beta=N/E_{tot}$. Firstly, extending 
some classical results by Kac~\cite{kac1959, ford1965statistical}, we focus on the empirical 
distribution of the momentum of one particle, computed from a unique long 
trajectory, namely
\begin{equation}
\label{eq:distr_emp_j}
\mathcal{P}_{e}^{(j)}\cbr{p}={1 \over T} \int_0^T\, dt\, \delta \cbr{p -p_j(t)} 
\,.
\end{equation}
Then we consider the marginal  probability distribution  
$\mathcal{P}_{e}\cbr{p,t}$ computed from the momenta $\{p_j\}$ of all the 
particles at a specific time $t$, i.e.
\begin{equation}
\label{eq:distr_emp_t}
\mathcal{P}_{e}\cbr{p,t}={1 \over N} \sum_{j=1}^N \delta \cbr{p -p_j(t)} \,.
\end{equation}
In both cases we assume that the system is prepared in an atypical initial 
condition. More precisely, we consider the case in which $Q_j(0)=0$, for all 
$j$, and  the total energy $E_{tot}$, at time $t=0$, is equally distributed 
among the momenta of the first $N^{\star}$ normal modes, with $1 \ll N^{\star} 
\ll N$:
\begin{equation}
P_j(0)=
\begin{cases}
\sqrt{2m E_{tot}/N^{\star}}\quad &\text{for}\quad 1 \le j \le N^{\star}\\
0\quad &\text{for}\quad N^{\star}< j \le N\,.
\end{cases}
\end{equation}
In this case, the dynamics of the first $N^{\star}$ normal modes is given by
\begin{equation}
\label{eq:dyn}
\begin{aligned}
Q(t)&=\sqrt{\frac{2 E_{tot}}{\omega_k^2N^{\star}}}\sin\cbr{\omega_k t}\\
P(t)&=\sqrt{\frac{2 m E_{tot}}{N^{\star}}}\cos\cbr{\omega_k t}\,.
\end{aligned}
\end{equation}

\subsection{Empirical distribution of single-particle momentum}
Our aim  is  to compute the empirical distribution of the momentum of a given 
particle $p_j$, i.e., the distribution of its values measured in time. This 
analytical calculation was carried out rigorously by Mazur and Montroll in 
Ref.~\cite{mazur1960poincare}. Here, we provide an alternative argument that has the advantage of being more concise and intuitive, in contrast to the mathematical rigour of~\cite{mazur1960poincare}. Our approach exploits the computation of 
the moments of the distribution; by showing that they are the same, in the limit 
of infinite measurement time, as those of a Gaussian, it is possible to conclude 
that the considered momentum follows the equilibrium Maxwell-Boltzmann 
distribution. The assumption $N\gg1$ will enter explicitly the calculation.

The momentum of the $j$-th particle can be written as a linear combination of 
the momenta of the normal modes by inverting Eq.~\eqref{eq:modes_p}:
\begin{equation}
\label{eq:mom}
\begin{aligned}
    p_j(t)&=\sqrt{\frac{2}{N+1}}\sum_{k=1}^N\, \sin\cbr{\frac{j k \pi}{N+1}} 
P_k(t)\\
&=2\sqrt{\frac{m 
E_{tot}}{(N+1)N^{\star}}}\sum_{k=1}^{N^{\star}}\sin\cbr{\frac{kj\pi}{N+1}}
\cos\cbr{\omega_k t}
\end{aligned}
\end{equation}
where the $\omega_k$'s are defined by Eq.~\eqref{eq:omega_m}, and the 
dynamics~\eqref{eq:dyn} has been taken into account. The $n$-th empirical moment 
of the distribution is defined as the average $\overline{p_j^n}$ of the $n$-th 
powerof $p_j$ over a measurement time $T$:
\begin{equation}
\label{eq:avp}
\begin{aligned}
    \overline{p_j^n}&=\frac{1}{T}\int_0^{T}dt p_j^n(t)\\
    &=\frac{1}{T}\int_0^{T}dt\,(C_{N^{\star}})^n 
\prod_{l=1}^n\sbr{\sum_{k_l=1}^{N^{\star}}\sin\cbr{\frac{k_l 
j\pi}{N+1}}\cos\cbr{\omega_{k_l} t}}\\
     &=(C_{N^{\star}})^n 
\sum_{k_1=1}^{N^{\star}}\dots\sum_{k_n=1}^{N^{\star}}\sin\cbr{\frac{k_1j\pi}{N+1
}}\dots\sin\cbr{\frac{k_nj\pi}{N+1}}\, 
\frac{1}{T}\int_0^{T}dt\,\cos\cbr{\omega_{k_1} t}\dots\cos\cbr{\omega_{k_n} t}
\end{aligned}
\end{equation}
with
\begin{equation}
    C_{N^{\star}}=2\sqrt{\frac{m E_{tot}}{(N+1)N^{\star}}}\,.
\end{equation}
We want to study the integral appearing in the last term of the above equation. 
To this end it is useful to recall that
\begin{equation}
    \frac{1}{2 \pi}\int_{0}^{2\pi}d \theta \cos^n(\theta)=
    \begin{cases}
    \frac{(n-1)!!}{n!!}\quad &\text{for $n$ even}\\
    0\quad &\text{for $n$ odd}\,.
    \end{cases}
\end{equation}
As a consequence, one has
\begin{equation}
\label{eq:avcos}
   \frac{1}{T}\int_{0}^{T}d t \cos^n(\omega t)\simeq
    \begin{cases}
    \frac{(n-1)!!}{n!!}\quad &\text{for $n$ even}\\
    0\quad &\text{for $n$ odd}\,.
    \end{cases}
\end{equation}
Indeed, we are just averaging over $\simeq \omega T/2 \pi$ periods of the 
integrated function, obtaining the same result we get for a single period, 
with a correction of the order $O\cbr{(\omega T)^{-1}}$. This correction comes from the fact that $T$ 
is not, in general, an exact multiple of $2 \pi/\omega$. 
If $\omega_1$, $\omega_2$, ..., $\omega_q$ are incommensurable (i.e., their 
ratios cannot be expressed as rational numbers), provided that $T$ is much 
larger than $(\omega_j-\omega_k)^{-1}$ for each choice of $1 \le k < j \le q$, a 
well known result~\cite{kac1959} assures that
\begin{equation}
\label{eq:mediacoseni}
\begin{aligned}
    \frac{1}{T}\int_{0}^{T}d t \cos^{n_1}(\omega_1 t)\cdot...\cdot 
\cos^{n_q}(\omega_q t) \simeq& \cbr{\frac{1}{T}\int_{0}^{T}d t 
\cos^{n_1}(\omega_1 t)}\cdot...\cdot\cbr{\frac{1}{T}\int_{0}^{T}d t 
\cos^{n_q}(\omega_1 t)}\\
    \simeq& \frac{(n_1-1)!!}{n_1!!}\cdot ...\cdot 
\frac{(n_q-1)!!}{n_q!!}\,\quad\text{if all $n$'s are even\,,}
\end{aligned}
\end{equation}
where the last step is a consequence of Eq.~\eqref{eq:avcos}. Instead, if at 
least one of the $n$'s is odd, the above quantity vanishes, again with 
corrections  due to the finite time $T$. Since the smallest sfrequency is 
$\omega_1$, one has that the error is at most of the order $O\cbr{q(\omega_1 
T)^{-1}}\simeq O(qN /\omega_0 T)$.

\begin{figure}[t!]
 \centering
 \includegraphics[width=0.9\linewidth]{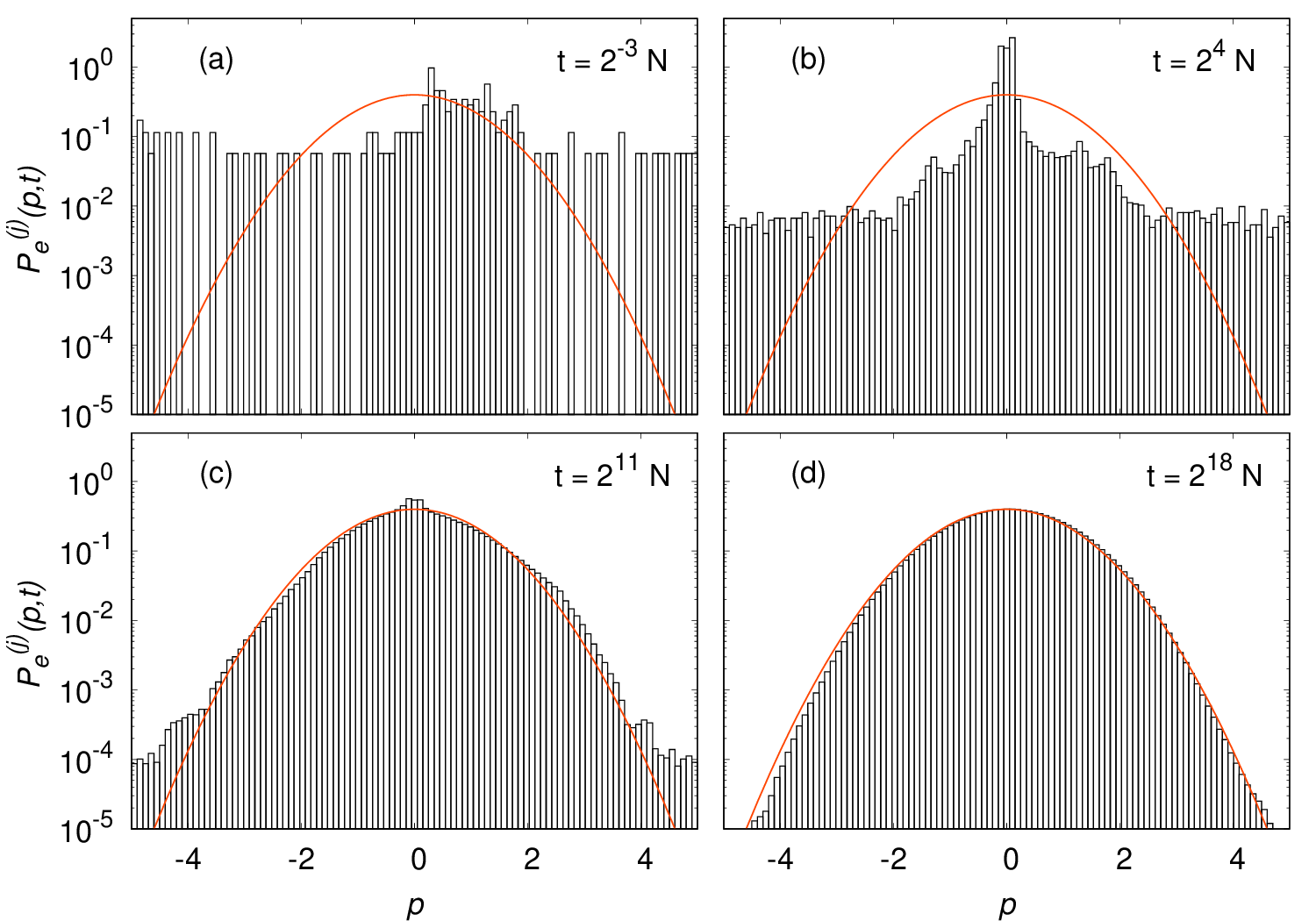}
 \caption{Convergence of the empirical distribution of the single particle 
momentum. The four panels show histograms for the momentum of the $j$-th 
particle, computed by considering the value of $p_j$ at integral times 
$0,1,2,...,t$. The system is prepared at time 0 in the atypical initial 
condition described in the text, where only $N^{\star}$ normal modes are 
excited. Different values of $t$ are considered: when the total time is large 
enough, the distribution approcahes the Maxwell-Boltzmann equilibrium (red 
curve).  Here $N=10^3$, $N^{\star}=0.1 N$, $E_{tot}=N$, $\omega_0=1$, $j=123$.  
\label{fig:1}}
\end{figure}

Let us consider again the integral in the last term of Eq.~\eqref{eq:avp}. The 
$\omega_k$'s are, in general, incommensurable. Therefore, the integral vanishes 
when $n$ is odd, since in that case at least one of the $\{n_l\}$, $l=1,...,q$, 
will be odd. When $n$ is even, the considered quantity is different from zero 
as soon as the $k$'s are pairwise equal, so that $n_1=...=n_q=2$. In the following we 
will neglect the contribution of terms containing groups of four or more equal 
$k$'s: if $n\ll N^{\star}$, the number of these terms is indeed $\sim 
O(N^{\star})$ times less numerous than the pairings, and it can be neglected if 
$N^{\star}\gg1$ (which is one of our assumptions on the initial condition). 
Calling $\Omega_n$ the set of possible pairings for the vector 
$\mathbf{k}=(k_1,...,k_l)$, we have then
\begin{equation}
    \overline{p_j^n}\simeq \cbr{\frac{C_{N^{\star}}}{\sqrt{2}}}^n \, 
\sum_{\mathbf{k}\in\Omega_n}\prod_{l=1}^n \sin\cbr{\frac{k_lj\pi}{N+1}}\,,
\end{equation}
with an error of $O(1/N^{\star})$ due to neglecting groups of 4, 6 and so on, 
and an error $O(nN/\omega_0 T)$ due to the finite averaging time $T$, as 
discussed before. Factor $2^{-n/2}$ comes from the explicit evaluation of 
Eq.~\eqref{eq:mediacoseni}\,.

At fixed $j$, we need now to estimate the sums appearing in the above equation, 
recalling that the $k$'s are pairwise equal. If $j> \frac{N}{N^{\star}}$, the 
arguments of the periodic functions can be thought as if independently extracted 
from a uniform distribution $\mathcal{P}(k)=1/N^{\star}$. One has:
\begin{equation}
 \av{\sin^2 \cbr{\frac{kj \pi}{N+1}}} \simeq 
\sum_{k=1}^{N^{\star}}\frac{1}{N^{\star}}\sin^2 \cbr{\frac{kj \pi}{N+1}} \simeq 
\frac{1}{2 \pi}\int_{-\pi}^{\pi}d \theta\, \sin^2(\theta)=\frac{1}{2}\,,
\end{equation}
and 
\begin{equation}
 \av{\prod_{l=1}^n \sin\cbr{\frac{k_lj\pi}{N+1}}} \simeq 2^{-n/2}\,,
\end{equation}
if $\mathbf{k}\in \Omega_n$.
As a consequence
\begin{equation}
\begin{aligned}
    \overline{p_j^n}&\simeq \cbr{\frac{C_{N^{\star}}}{2}}^n (N^{\star})^{n/2} \, 
\mathcal{N}(\Omega_n)\simeq\cbr{\frac{m 
E_{tot}}{N+1}}^{n/2}\mathcal{N}(\Omega_n)\,,
\end{aligned}
\end{equation}
where $\mathcal{N}(\Omega_n)$ is the number of ways in which we can choose the 
pairings. These are the moments of a Gaussian distribution with zero average and 
$\frac{m E_{tot}}{N+1}$ variance. 

Summarising, it is possible to show that, if $n \ll N^{\star} \ll N$, the first 
$n$ moments of the distribution are those of a Maxwell-Boltzmann distribution. 
In the limit of $N\gg1$ with $N^{\star}/N$ fixed, the Gaussian distribution is 
thus recovered up to an arbitrary number of moments. Let us note that the 
assumption $Q_j(0)=0$, while allowing to make the calculations clearer, is not 
really relevant. Indeed, if $Q_j(0)\neq 0$ we can repeat the above computation 
while replacing $\omega_k t$ by $\omega_k t + \phi_k$, where the phases $\phi_k$ 
take into account the initial conditions. 

Fig.~\ref{fig:1} shows the standardized histogram of the relative 
frequencies of single-particle velocities of the considered system, in the $N 
\gg 1$ limit, with the initial conditions discussed before. As expected, the 
shape of the distribution tends to a Gaussian in the large-time limit.

\subsection{Distribution of momenta at a given time}
A similar strategy can be used to show that, at any given time $t$ large enough, 
the histogram of the momenta is well approximated by a Gaussian distribution. 
Again, the large number of degrees of freedom plays an important role.
We want to compute the empirical moments
\begin{equation}
    \av{p^n}(t)=\frac{1}{N}\sum_{j=1}^N p_j^n(t)\,,
\end{equation}
defined according to the distribution $\mathcal{P}_{e}^{(j)}\cbr{p}$ introduced 
by Eq.~\eqref{eq:distr_emp_j}.
Using again Eq.~\eqref{eq:mom} we get
\begin{equation}
    \begin{aligned}
\av{p^n}(t)=&\frac{1}{N}\sum_{j=1}^{N}(C_{N^{\star}})^{n}\sbr{\sum_{k=1}^{N^{
\star}}\sin\cbr{\frac{kj\pi}{N+1}}\cos\cbr{\omega_k t}}^{n}\\    
=&\frac{1}{N}(C_{N^{\star}})^{n}\sum_{k_1}^{N^{\star}}\dots\sum_{k_{n}=1}^{N^{
\star}} 
\sbr{\prod_{l=1}^{N}\cos\cbr{\omega_{k_l}t}}\sum_{j=1}^N\sin\cbr{\frac{k_1 j 
\pi}{N+1}}\dots \sin\cbr{\frac{k_{n} j \pi}{N+1}}\,.
    \end{aligned}
\end{equation}
Reasoning as before, we see that the sum over $j$ vanishes in the large $N$ 
limit unless the $k$'s are pairwise equal. Again, we neglect the terms including 
groups of 4 or more equal $k$'s, assuming that $n\ll N^{\star}$, so that their 
relative contribution is $O(1/N^{\star})$. That sum selects paired values of $k$ 
for the product inside the square brackets, and we end with
\begin{equation}
\av{p^n}(t)\simeq\frac{1}{N}(C_{N^{\star}})^{n}\sum_{\mathbf{k}\in\Omega_n}\sbr{
\prod_{l=1}^{N}\cos\cbr{\omega_{k_l}t}}\,.
\end{equation}
If $t$ is ``large enough'' (we will come back to this point in the following 
section), different values of $\omega_{k_l}$ lead to completely uncorrelated 
values of $\cos(\omega_{k_l} t)$. Hence, as before, we can consider the 
arguments of the cosines as extracted from a uniform distribution, obtaining
\begin{equation}
    \av{p^n}(t)\simeq \cbr{\frac{C_{N^{\star}}}{2}}^n (N^{\star})^{n/2} \, 
\mathcal{N}(\Omega_n)\simeq\cbr{\frac{m 
E_{tot}}{N+1}}^{n/2}\mathcal{N}(\Omega_n)\,.
\end{equation}
These are again the moments of the equilibrium Maxwell-Boltzmann distribution. 
We had to assume $n \ll N^{\star}$, meaning that a Gaussian distribution is recovered 
only in the limit of large number of degrees of freedom.

The 
empirical distribution can be compared with the Maxwell-Boltzmann by looking at 
the Kullback-Leibler divergence $K(\mathcal{P}_e(p,t), \mathcal{P}_{MB}(p))$ 
which provides a sort of distance between the empirical $\mathcal{P}_e(p,t)$ and 
the Maxwell-Boltzmann:
\begin{equation}
\label{eq:kl}
   K[\mathcal{P}_e(p,t), \mathcal{P}_{MB}(p)]= - \int  \mathcal{P}_e(p,t) \ln 
\frac{ \mathcal{P}_{MB}(p)}{\mathcal{P}_e(p,t)} dp.
\end{equation}

\begin{figure}[t!]
 \centering
 \includegraphics[width=0.9\linewidth]{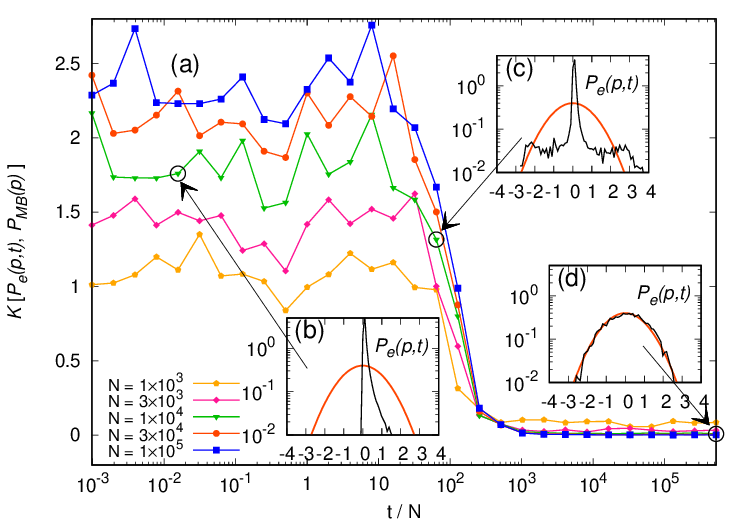}
 \caption{Kullback-Leibler divergence between the empirical distribution of the 
momenta~\eqref{eq:distr_emp_t} and the Maxwell-Boltzmann~\eqref{eq:mb}, as a 
function of time. Panel (a) shows the behaviour of $K[\mathcal{P}_e(p,t), 
\mathcal{P}_{MB}(p)]$, defined by Eq.~\eqref{eq:kl}, for different values of 
$N$. As before, the system is initialized at time $t=0$ in a 
far-from-equilibrium state where only a small fraction of the modes is excited. The transition to equilibrium takes place on a time scale that is proportional to $N$, as expected from the argument in the main text.
Panels (b)-(d) show the detail of the empirical distribution 
$\mathcal{P}_e(p,t)$ at different times, for the $N=10^4$ case. Unlike in 
Fig.~\ref{fig:1}, here the histogram is built including the momenta of 
\textit{all} particles in the system, at the considered time. Parameters: 
$N^{\star}=0.1N$, $E_{tot}=N$, $\omega_0=1$. 
 \label{fig:2}}
\end{figure}

Figure~\ref{fig:2} shows how the Kullback-Leibler divergences 
approach their equilibrium limit, for different values of $N$. As expected, the transition happens on a time scale that depends linearly on $N$.

A comment is in order: even if this behaviour 
may look similar to the H-Theorem for diluited gases, such a resemblance is 
only superficial. Indeed, while in the cases of diluited gases the approach to 
the Maxwell-Boltzmann is due to the collisions among different particles that actually exchange 
energy and momentum, in the considered case the ``thermalization'' is due to a dephasing 
mechanism.

\section{Analysis of the time scales}
\label{sec::times}

In the previous section, when considering the distribution of the momenta at a 
given time, we had to assume that $t$ was ``large enough'' in order for our 
approximations to hold. In particular we required $\cos(\omega_{k_1}t)$ and 
$\cos(\omega_{k_2}t)$ to be uncorrelated as soon as $k_1 \ne k_2$. Such a dephasing 
hypothesis amounts to asking that
\begin{equation}
|\omega_{k_1}t-\omega_{k_2}t|> 2\pi c\,,
\end{equation}
where $c$ is the number of phases by which the two oscillator have to differ before
they can be considered uncorrelated. The constant $c$ may be much larger than 1,
but it is not expected to depend strongly on the size $N$ of the system. In other words,
we require
\begin{equation}
t> \frac{c}{|\omega_{k_1}-\omega_{k_2}|}
\end{equation}
for each choice of $k_1$ and $k_2$. To estimate this typical relaxation time, we 
need to pick the minimum value of $|\omega_{k_1}-\omega_{k_2}|$ among the 
possible pairs $(k_1,k_2)$. This term is minimized when $k_1=\tilde{k}$ and 
$k_2=\tilde{k}-1$ (or vice-versa), with $\tilde{k}$ chosen such that 
$\omega_{\tilde{k}}-\omega_{\tilde{k}-1}$
is minimum. In the large-$N$ limit this quantity is approximated by
\begin{equation}
\omega_{\tilde{k}}-\omega_{\tilde{k}-1}=\omega_0\sin\cbr{\frac{\tilde{k} 
\pi}{2N+2}}-\omega_0\sin\cbr{\frac{\tilde{k} \pi- \pi}{2N+2}}\simeq 
\omega_0\cos\cbr{\frac{\tilde{k} \pi}{2N+2}}\frac{\pi}{2N+2}\,,
\end{equation}
which is minimum when $\tilde{k}$ is maximum, i.e. for $\tilde{k}=N^{\star}$.

Dephasing is thus expected to occur at
\begin{equation}
t> \frac{4cN}{\omega_0\cos\cbr{\frac{N^{\star} \pi}{2N}}}\,,
\end{equation}
i.e. $t>4cN/\omega_0$ in the $N^{\star}/N \ll 1$ limit.

It is instructive to compare this characteristic time with the typical 
relaxation time of the ``damped'' version of the considered system. For doing so, we assume that our chain of oscillators is now in contact with a viscous medium which acts at the same time as a thermal bath and as a source of viscous friction. By 
considering the (stochastic) effect of the medium, one gets the Klein-Kramers 
stochastic process~\cite{gardiner1985handbook, marino2016advective}
\begin{equation}
\begin{split}
    & \frac{\partial q_j}{\partial t}=\frac{p_j}{m} \\
    & \frac{\partial p_j}{\partial t}=\omega_0^2(q_{j+1} - 2 q_{j} + q_{j-1}) 
-\gamma p_j + \sqrt{2 \gamma T} \xi_j  \,
\end{split}
    \label{eq:stoch_case}
\end{equation}
where $\gamma$ is the damping coefficient and $T$ is the temperature of the 
thermal bath (we are taking the Boltzmann constant $k_B$ equal to 1). Here the 
$\{\xi_j\}$ are time-dependent, delta-correlated Gaussian noises such that 
$\av{\xi_j(t)\xi_k(t')}=\delta_{jk}\delta(t-t')$.
Such  a system is surely ergodic and the stationary probability distribution is 
the familiar equilibrium one
\begin{equation}
\mathcal{P}_{s}(\mathbf{q},\mathbf{p}) \propto 
e^{-\frac{H(\mathbf{q},\mathbf{p})}{T}}.
\end{equation}

Also in this case we can consider the evolution of the normal modes. By taking 
into account Eqs.~\eqref{eq:modes} and~\eqref{eq:stoch_case} one gets
\begin{equation}
\begin{aligned}
\dot{Q_k}&=\frac{1}{m} P_k\\
\dot{P_k}&=- \omega_k^2 Q_k - \frac{\gamma}{m} P + \sqrt{2 \gamma T} \zeta_k
\end{aligned}
\end{equation}
where the $\{\zeta_k\}$ are again delta-correlated Gaussian noises. It is 
important to notice that also in this case the motion of the modes is 
independent (i.e. the friction does not couple normal modes with  different 
$k$); nonetheless, the system is ergodic, because the presence of the noise 
allows it to explore, in principle, any point of the phase-space.

The Fokker-Planck equation for the evolution of the probability density function 
$\mathcal{P}\cbr{Q_k,P_k,t}$ of the $k$-th normal mode can be derived using 
standard methods~\cite{gardiner1985handbook}:
\begin{equation}
\label{eq:fokkerplanck}
\partial_t\mathcal{P}=-\partial_{Q_k} \cbr{P_k\mathcal{P}}+\partial_{P_k} 
\cbr{\omega_k^ 2Q_k\mathcal{P}+\frac{\gamma}{m}P_k\mathcal{P}}+\gamma 
T\partial_{P_k}^2 \mathcal{P}\,.
\end{equation}

The above equation allows to compute also the time dependence of the correlation 
functions of the system in the stationary state. In particular one gets
\begin{equation}
\frac{d}{dt}\av{Q_k(t) Q_k(0)}=\frac{1}{m}\av{P_k(t)Q_k(0)}
\end{equation}
and 
\begin{equation}
\frac{d}{dt}\av{P_k(t) Q_k(0)}-\omega_k^2 m \av{Q_k(t) Q_k(0)} 
-\frac{\gamma}{m}\av{P_k(t) Q_k(0)}\,,
\end{equation}
which, once combined together, lead to
\begin{equation}
\frac{d^2}{d t^2}\av{Q_k(t) Q_k(0)}+\frac{\gamma}{m}\frac{d}{dt}\av{Q_k(t) 
Q_k(0)}+ \omega_k^2\av{Q_k(t) Q_k(0)}=0\,.
\end{equation}
For $\omega_k <\gamma/m$ the solution of this equation admits two characteristic 
frequencies $\tilde{\omega}_{\pm}$, namely
\begin{equation}
\tilde{\omega}_{\pm}=\frac{\gamma}{2m}\cbr{1 \pm \sqrt{1-\frac{m^2 
\omega_k^2}{\gamma^2}}}.
\end{equation}
In the limit $\omega_k \ll \gamma/m$ one has therefore
\begin{equation}
\tilde{\omega}_- \simeq \frac{m}{4 \gamma} \omega_k^2 \simeq \frac{m \omega_0^2 
\pi^2 k^2}{\gamma N^2}\,.
\end{equation}

Therefore, as a matter of fact, even in the damped case the system needs a time 
that scales as $N^2$ in order to get complete relaxation for the modes. As we 
discussed before, the dephasing mechanism that guarantees for ``practical'' 
ergodicity in the deterministic version is instead expected to occur on time scales of order $O(N)$.

\section{Conclusions}
\label{sec::conclusion}
The main aim of this paper was to expose, at a pedagogical level, some aspects
of the foundation of statistical mechanics, namely the role of 
ergodicity for the validity of the statistical approach to the study of complex 
systems.  

 We analyzed a chain of classical harmonic oscillators (i.e. a paradigmatic 
example of integrable system, which cannot be suspected to show chaotic 
behaviour). By extending some well-known results by Kac~\cite{kac1959}, we showed that the 
Maxwell-Bolzmann distribution approximates with arbitrary precision (in the 
limit of large number of degrees of freedom) the empirical distribution of the 
momenta of the system, after a dephasing time which scales with the size of the 
chain. This is true also for quite pathological initial conditions, where only a 
small fraction of the normal modes is excited at time $t=0$. The scaling of the 
typical dephasing time with the number of oscillators $N$ may appear as a limit 
of our argument, since this time will diverge in the thermodynamic limit; on the 
other hand one should consider, as explicitely shown before, that the damped 
version of this model (which is ergodic by definition) needs times of the order 
$O(N^2)$ to reach thermalization for each normal mode. 
 
 This comparison clearly shows that the effective thermalization observed in 
large systems has little to do with the mathematical concept of ergodicity, and 
it is instead related to the large number of components concurring to define the 
global observales that are usually taken into account (in our case, the large 
number of normal modes that define the momentum of a single particle). When 
these components cease to be in phase, the predictions of statistical mechanics 
start to be effective; this can be observed even in integrable systems, without 
need for the mathematical notion of ergodicity to hold.

In other words, we believe that the present work give further evidence of the 
idea (which had been substantiated mathematically by  Khinchin, Mazur and  van 
der Linden) that the  most relevant  ingredient of statistical mechanics is the 
large number of degrees of freedom, and the global nature of the observables 
that are typically taken into account.

\section*{Acknowledgements}

 RM is supported by \#NEXTGENERATIONEU (NGEU) and funded by the Ministry of University and Research (MUR), National Recovery and Resilience Plan (NRRP), project MNESYS (PE0000006) "A Multiscale integrated approach to the study of the nervous system in health and disease" (DN. 1553 11.10.2022).



\bibliography{cas-refs}





\end{document}